\documentclass[preprint2]{aastex}

\shorttitle{{\it BeppoSAX} Observation of the NGC 3079 Nucleus}
\shortauthors{Iyomoto et al.}

\begin{document}

\title{{\it BeppoSAX} Observation of NGC 3079}
\author{Naoko Iyomoto\altaffilmark{1}, Yasushi Fukazawa\altaffilmark{2},
Naomasa Nakai\altaffilmark{3,4}, Yuko Ishihara\altaffilmark{3,4}}

\altaffiltext{1}{Institute of Space and Astronautical Science,
		Yoshinodai, Sagamihara, Kanagawa, 229-8510, Japan}
\altaffiltext{2}{Department of Physical Science, Graduate School of Science, 
		Hiroshima University, 1-3-1 Kagamiyama, Higashi-Hiroshima, 
		Hiroshima, 739-8526, Japan}
\altaffiltext{3}{Nobeyama Radio Observatory, 
		Minamimaki-mura, Minamisaku-gun, Nagano, 384-1305, Japan}
\altaffiltext{4}{Department of Astronomy, School of Science, 
		University of Tokyo,
		7-3-1 Hongo, Bunkyo-ku, Tokyo, 113-0033, Japan}

\begin{abstract}
Using the {\it BeppoSAX} observatory,
we have observed a nearby LINER/Seyfert 2 galaxy, NGC 3079,
which is known as an outflow galaxy and a bright H$_2$O-maser source.
Using the PDS detector,
we have revealed that the NGC 3079 nucleus suffers from 
a Compton-thick absorption,
with a hydrogen column density $\sim 10^{25}$ cm$^{-2}$.
After corrected the absorption, 
the 2--10 keV luminosity becomes
$10^{42-43}$ erg s$^{-1}$
at a distance of 16 Mpc.
It is 2--3 orders of magnitude higher than
that observed in the MECS band (below 10 kev).
We also detected a strong Fe-K line at $6.4^{+0.3}_{-0.2}$ keV
with an equivalent width of $2.4^{+2.9}_{-1.5}$ keV,
which is consistent with the heavy absorption.
\end{abstract}


\section{Introduction}

NGC 3079 is an edge-on spiral galaxy at our vicinity 
(16 Mpc at $H_0 = 75$ km s$^{-1}$ Mpc$^{-1}$).
It exhibits AGN and/or starburst activities in various wave band.
It is known as an outflow galaxy in radio (Duric \& Seaquist 1988), 
H$_\alpha$ (Veilleux et al. 1994) and X-ray (Pietsch et al. 1998).
In optical, it is classified as 
a LINER (low-ionization nuclear emission region) or a Seyfert 2,
but whether the H$_{\alpha}$ line is broad or not 
is in controversy (Ho et al. 1997).
In far infrared, it is luminous and shows an extended emission,
indicating a dominant contribution of the starburst activity 
(Perez Garcia et al. 2000).
In radio, 
it is a bright H$_2$O-maser source (Nakai et al. 1995;
Trotter et al. 1998; Sawada-Satoh et al. 2000).

In X-ray, 
using the {\it ROSAT} and {\it ASCA} observatories,
Dahlem et al. (1998) and Ptak et al. (1999)
reported that the NGC 3079 emission is dominated 
by a thin-thermal component and a power-law component 
below and above $\sim 2$ keV, respectively.
The former is attributed to the hot gas in the host galaxy.
The latter 
with 0.4--10 keV luminosity of $\sim 3 \times 10^{40}$ erg s$^{-1}$
can be attributed to an assembly of X-ray binaries.
Super-nova remnants and hot gas due to the starburst activity
can also contribute to the latter component.
Therefore, 
there is little room for the AGN emission in the {\it ASCA} spectrum.
Due to poor photon statistics,
the presence of Fe-K emission line have not been constrained with {\it ASCA}.

We inferred that 
the faintness of the NGC 3079 nucleus in X-ray is due to complete obscuration,
and performed a {\it BeppoSAX} observation of NGC 3079.

\section{Observation and Data reduction}
We observed NGC 3079 with {\it BeppoSAX} on 26--27 May 2000.
The exposure time of LECS, MECS2, MECS3 and PDS
are 13.6, 44.5, 44.5 and 21.0 ksec, respectively.
The HPGSPC was not operated.
Figure 1 shows the LECS and MECS2+3 images of NGC 3079.
The brightness peak coincides in position with the radio/optical nucleus.
We utilized the LECS and MECS2+3 spectra 
supplied by the {\it BeppoSAX} Science Data Center (SDC),
which are integrated $2'$ around the brightness peak.
We utilized the blank-sky files supplied by SDC
to subtract the LECS and MECS2+3 backgrounds.
We also utilized the PDS spectrum supplied by SDC. 
We rebinned the spectra to have grater than 15 counts per bin.
Net count rate of the LECS, MECS2+3 and PDS are 
$(2.9\pm 0.5)\times 10^{-3}$, $(4.5\pm 0.4)\times 10^{-3}$
and $(1.8\pm 0.3)\times 10^{-1}$ count s$^{-1}$,
respectively.

\section{Spectra}
Figure 2 shows the LECS, MECS2+3 and PDS spectra of NGC 3079.
It is clear that the data requires at least three components:
a featureless continuum in the LECS and MECS band, 
a heavily absorbed continuum in the PDS band, 
and a strong emission line in the MECS band. 
On the other hand,
the soft thin-thermal components
that detected in the {\it ROSAT} and {\it ASCA} spectra
are not statistically significant
in the {\it BeppoSAX} spectra,
due to poor statistics at low energy.
In this letter, we are not interested in the soft components, 
so that we deal here with the MECS and PDS spectra only above 2 keV,
where the soft component is negligible. 
Also, 
we will not examine intrinsic absorption for the MECS-band continuum.

\subsection{The MECS spectrum}
The MECS continuum above 2 keV is well represented
with a power-law having a photon index of $\sim 2$.
The 2--10 keV luminosity becomes $\sim 1 \times 10^{40}$ erg s$^{-1}$.
These are consistent with the {\it ASCA} results
within the model uncertainty.
The structure around 6 keV is well reproduced with a Gaussian.
Figure 3 shows confidence contours of the Gaussian component.
The line center energy of $6.4^{+0.3}_{-0.2}$ keV is
in good agreement with the K-fluorescent line from cold iron.
As shown in Model (a) in Table 1, 
the line width is not resolved with the MECS energy resolution.
Therefore, hereafter, we assume that the Fe-K line is narrow (Model b).
The equivalent width against the power-law component becomes $2.4^{+2.9}_{-1.5}$ keV in the 90\% confidence level.
It is consistent with the {\it ASCA} loose upper limit,
although no evidence of Fe-K line was observed in the {\it ASCA} spectra
(Dahlem et al. 1998; Ptak et al. 1999).
Such a large equivalent width cannot be observed
if the continuum emission from the nucleus is directly seen
(e.g. Makishima et al. 1986).
Consequently, 
the strong emission line suggests
that the direct emission 
is completely blocked.

\subsection{The PDS spectrum}
Indeed, 
the PDS spectrum exceeds two orders of magnitude over
the extrapolation of the best-fit Model of the MECS spectrum.
Althogh the PDS has a large beam size (1.4$^\circ$ FWHM),
we can reasonably conclude that
the contribution of contamination sources to the excess is small,
as follows.
In the MECS image (Figure 1b; 1$^\circ$ circular FOV), 
there are no sources bright enough to explain the PDS excess.
For example, 
the observed flux of the second brightest source in the MECS image, 
QSO 0957-5608, is $\sim$ 40\% of that of NGC 3079, in the MECS band.
Similarly, 
the sources in the ROSAT/PSPC image (2$^\circ$ circular FOV) are relatively faint.
The 90\% upper limits to the 20-100 keV flux
of QSO 0957-5608 in the MECS
and the other sources in the PSPC are estimated to be
$2\times 10^{-13}$ and $3\times 10^{-12}$ erg cm$^{-2}$ s$^{-1}$,
respectively.
These are two or one order of magnitudes fainter than 
the PDS flux in the 20--100 keV band.
Accordingly, 
the PDS excess is attributed to 
a heavily absorbed emission of the NGC 3079 nucleus,
unless there are serendipitous Compton-thick sources in the PDS field-of-view.

Therefore, 
we fitted the PDS spectrum with an absorbed power-law model.
Figure 4 shows confidence contours of the power-law component.
The photon index is consistent with those seen in Seyfert galaxies,
although it is not strictly constrained (Model c).
Therefore, hereafter, we fix the photon index at 1.9,
a canonical value of Seyfert galaxies (Model d).
With the fixed photon index,
a heavy absorption with a column density
larger than $3 \times 10^{24}$ cm$^{-2}$ is required.
After corrected the absorption,
the 2--10 keV and 20--100 keV luminosities become
$\sim 1.2$ and $\sim 1.5 \times 10^{42}$ erg s$^{-1}$, respectively.
These are two orders of magnitude higher than that observed in the MECS band.

In the spectral fitting,
we considered only the photoelectric absorption,
using the ``wabs'' model in XSPEC.
In such a Compton-thick environment, however, 
the Compton scattering becomes as important as the photoelectric absorption.
As investigated in Matt et al. (1999),
the ``wabs'' model underestimates the flux 
0.5--1 order of magnitude
in the column density range of $10^{24.5-25}$ cm$^{-2}$.
Considering the absorption due to the Compton scattering,
the 2--10 keV and 20--100 keV luminosities of NGC 3079
are estimated to be $10^{42-43}$ erg s$^{-1}$.

\subsection{Combined fit of the MECS and PDS spectra}
Using the three-component model,
we performed combined fits of the MECS and PDS spectra.
As shown in Table 1,
the results are similar to those of the individual fits,
except for an improvement of the lower limit to the absorption column density.
Model (f) in Table 1 shows our best result.
We superposed the best-fit model with each component
on the spectra in Figure 2.
The residual on the LECS band 
is roughly consistent with the soft component measured with {\it ROSAT} and {\it ASCA}.

Starting from the best-fit model,
we examined additional structures in the NGC 3079 spectrum.
First, we added the ``pexrav'' model (Magdziarz \& Zdziarski 1995) 
to the best-fit model, to study the Compton-reflection hump.
We fixed the photon index of the reflection component at 1.9.
The data prefers no reflection
with an upper limit to the reflection efficiency of 23\%.
It is similar to the situation seen in the other Compton-thick AGNs,
such as M51 (Fukazawa et al. 2001) and NGC 4945 (Guainazzi et al. 2000).
The small reflection efficiency indicates that
not only the reflection process but also the transmission process 
contributes to the strong Fe-K line of NGC 3079.
Secondly, we applied a high-energy exponential cutoff 
to the absorbed power-law component of the best-fit model.
However, the cutoff energy was not restricted well
with a loose lower limit of 16 keV.

\section{Variability}
In Figure 5,
we show the LECS, MECS2+3 and PDS short-term light curves of NGC 3079.
In the LECS and MECS band,
no short-term variability was detected during one-day observation.
It is consistent with the spectral result
that the direct component of the NGC 3079 nucleus is not observed below 10 keV.
In the PDS band, 
it is difficult to constrain the short-term variability,
due to the poor photon statistic.

To study long-term variability,
we fitted the {\it ASCA} spectra above 2 keV
in the same way as Model (b).
We fixed the photon index of the power-law component
at the best-fit value of the {\it BeppoSAX} data. 
We are not concerned with the absorption.
The 2--10 keV flux 
becomes $(5\pm 2) \times 10^{-13}$ erg cm$^{-2}$ s$^{-1}$,
so that no long-term variation was detected during seven years.
It is in contrast to 
the three-times long-term variation of the M51 nucleus (Fukazawa et al. 2001).
Therefore, it might suggest that 
the NGC 3079 emission below 10 keV is mainly due to the host galaxy emission 
rather than the scattered and/or reflected component of the nuclear emission.

\section{Discussion}
The main result of the {\it BeppoSAX} observation of NGC 3079 is 
the detection of a heavily absorbed continuum and a strong Fe-K emission line.

The column density of the absorbed continuum
and the equivalent width of the Fe-K line
allow us to estimate the geometry of the system.
Adopting the geometry in Ghisellini et al. (1994),
the obtained values of the column density and equivalent width
indicate a 
high inclination angle of 60--80 degree,
corresponding to the dominant contribution of 
the transmission component of the Fe-K emission line.
It is consistent with the no detection of the Compton hump
in the {\it BeppoSAX} spectra.

Using the absorption-corrected 2--10 keV luminosity, 
we can estimate 
how active the NGC 3079 nucleus is.
Trotter et al. (1998) estimated that 
the binding mass of the inner parsec of NGC 3079 is $\sim 10^6$ $M_\odot$
by the H$_2$O maser kinematics.
Accordingly, the X-ray emission of NGC 3079 is 
1--10\% of the Eddington luminosity.
It is as active as the Seyfert galaxies and quasars.

We can also examine 
whether the AGN or starburst activity dominates the NGC 3079 energy output
by comparing the 2--10 keV luminosity
with the infrared luminosity.
The infrared luminosity of NGC 3079 becomes
$1.5\times 10^{44}$ erg s$^{-1}$ (the NED data base).
Assuming that 
the 2--10 keV luminosity is $\sim 3$\% of the bolometric luminosity,
as typical for quasars (Elvis et al. 1994),
the bolometric luminosity associated to the AGN 
is estimated to be as luminous as the observed infrared luminosity,
hence the AGN and starburst activities 
contribute to the energy output of NGC 3079 in the same order of magnitude.

\begin{figure}
\epsscale{1.0}
\plotone{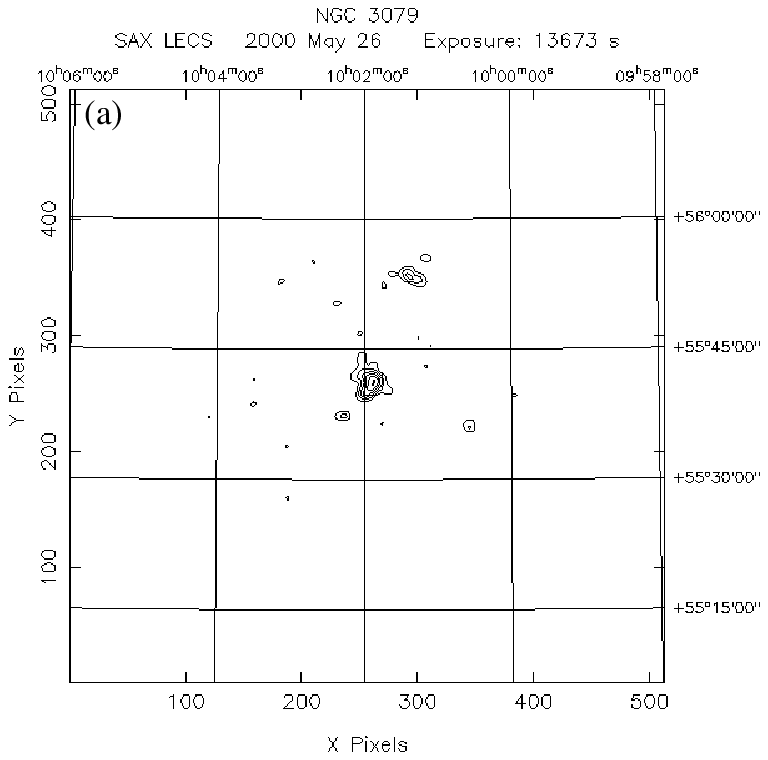}\\
\bigskip
\plotone{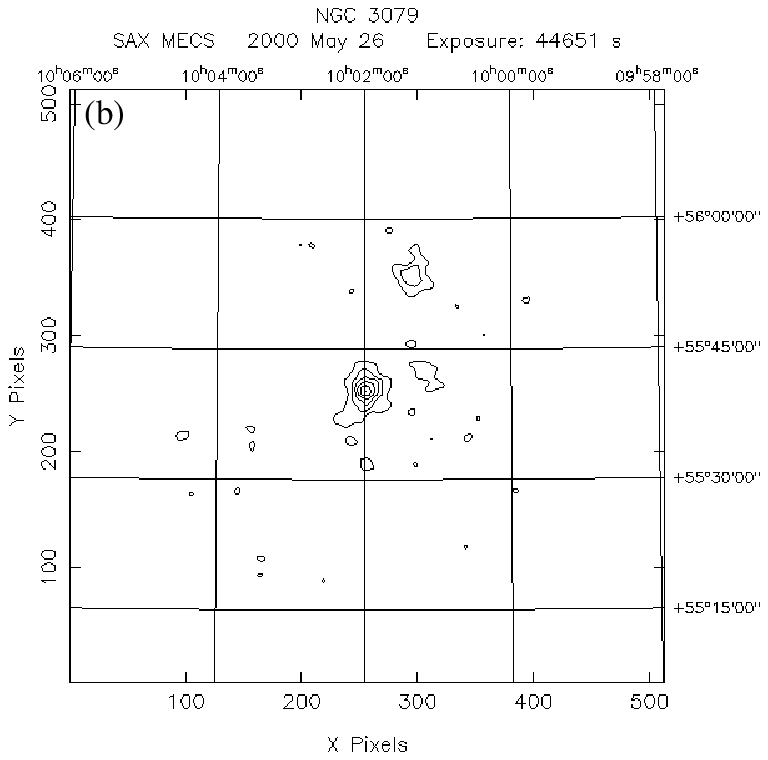}
\caption{
(a) The LECS image supplied by the {\it BeppoSAX} Science Data Center.
Shown after Gaussian smoothing with $0'.5$ sigma.
Contours correspond to 0.04, 0.08, 0.12 and 0.16 counts/pixel.
(b) Same as (a), but for the MECS2+3.
Contours correspond to 0.2, 0.3, 0.4, 0.5, 0.6 and 0.7 counts/pixel.
The brightest peak coincide in position with NGC 3079.
The second brightest peak 12$'$ north-northwest of NGC 3079 
is a gravitationally-lensed quasar, QSO 0957-5608.
\label{fig1}
}
\end{figure}

\begin{figure}
\epsscale{1.0}
\plotone{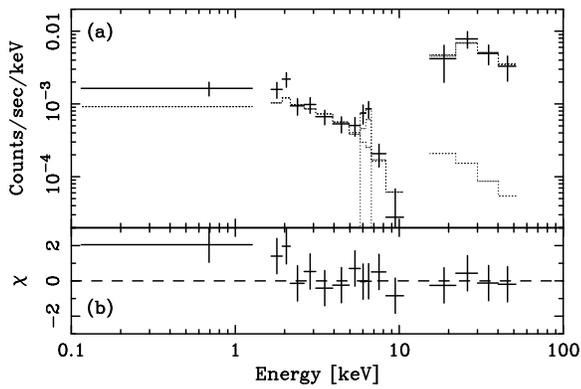}
\caption{
(a) Crosses show the LECS, MECS2+3 and PDS spectra of NGC 3079.
The best-fit model above 2 keV (Model f)
with each model component is superposed as histograms.
(b) Residual of the best-fit model.
Residuals below 2 keV may be attributed to the soft components
observed in the {\it ROSAT} and {\it ASCA} spectra.
\label{fig2}
}
\end{figure}

\begin{figure}
\epsscale{1.0}
\plotone{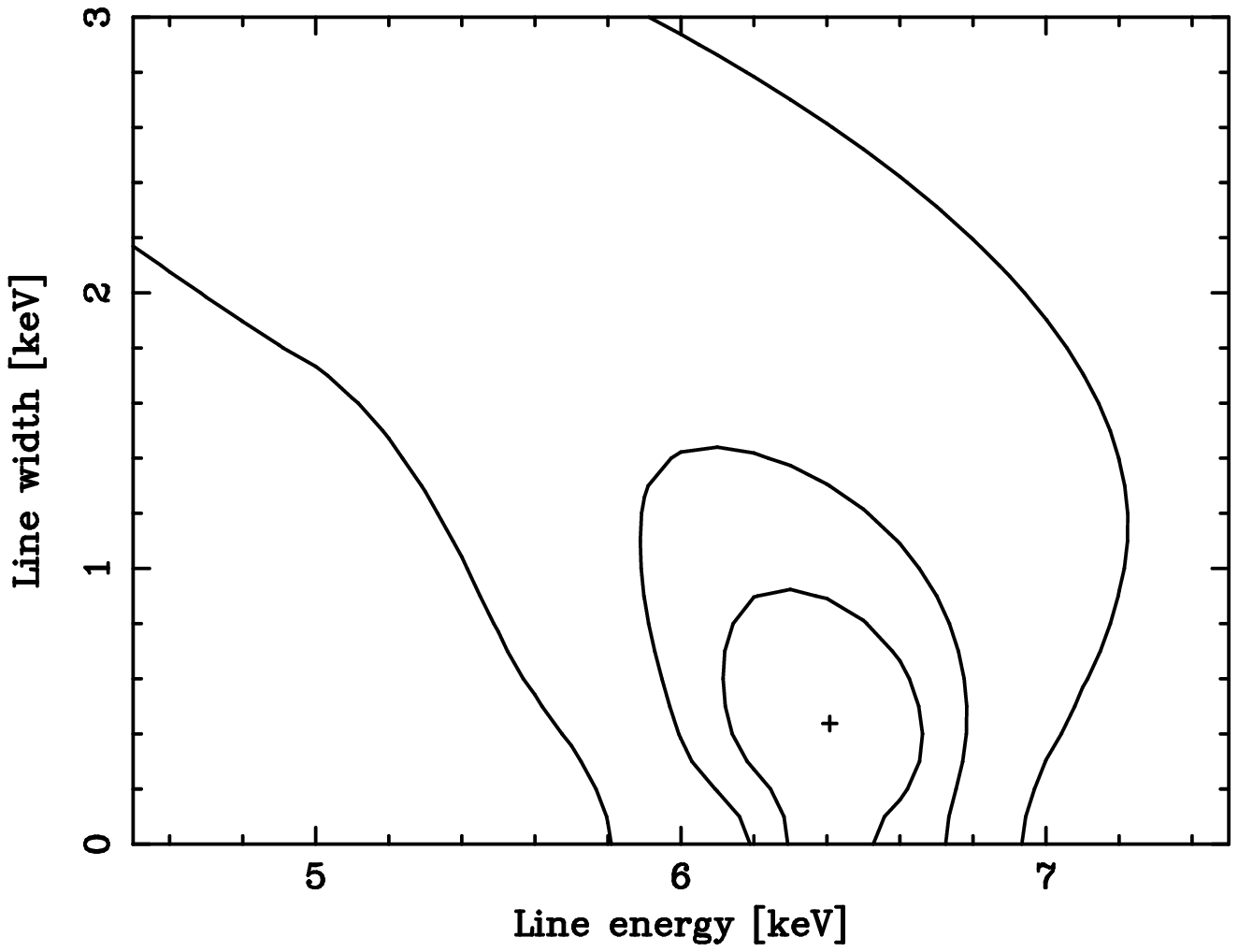}\\
\plotone{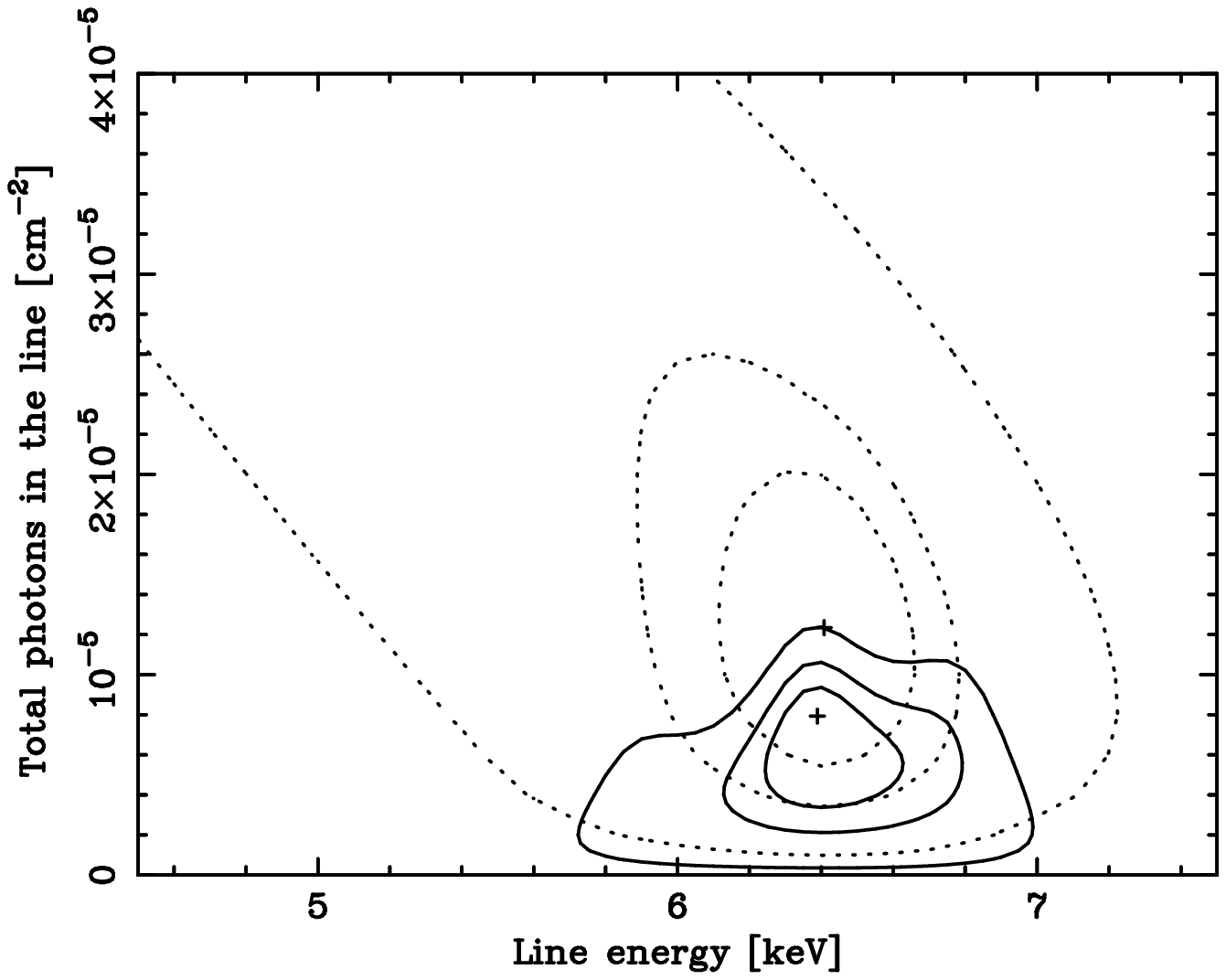}
\caption{
Two-parameter 68\%, 90\% and 99\% confidence contours of the Gaussian component.
(a) Confidence contours of the center energy and line width of the Fe-K line (Model a).
(b) Confidence contours of the center energy and normalization.
The dotted and solid contours correspond to
the free line width (Model a)
and that fixed at 0 eV (Model b), respectively.
\label{fig3}
}
\end{figure}

\begin{figure}
\epsscale{1.0}
\plotone{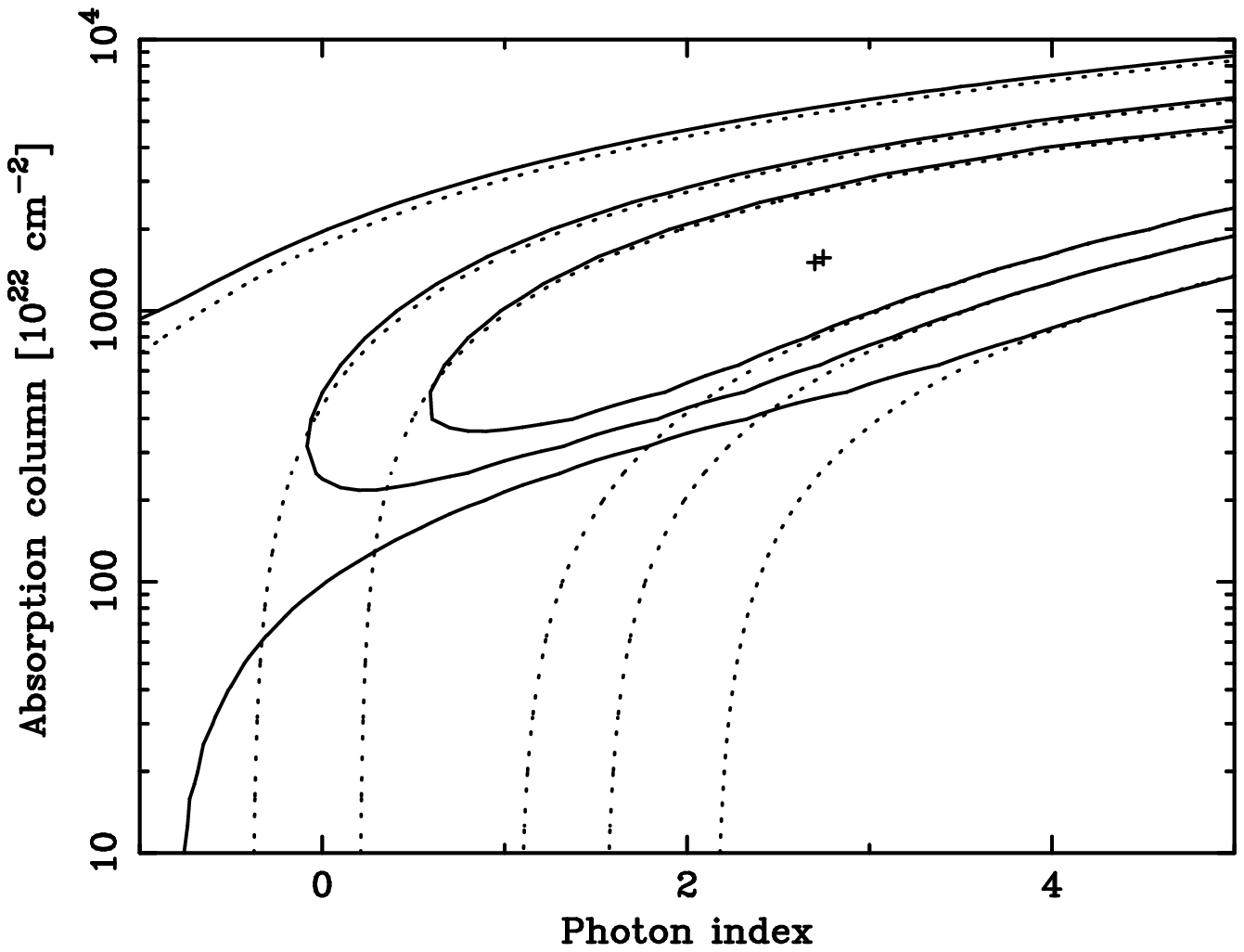}\\
\plotone{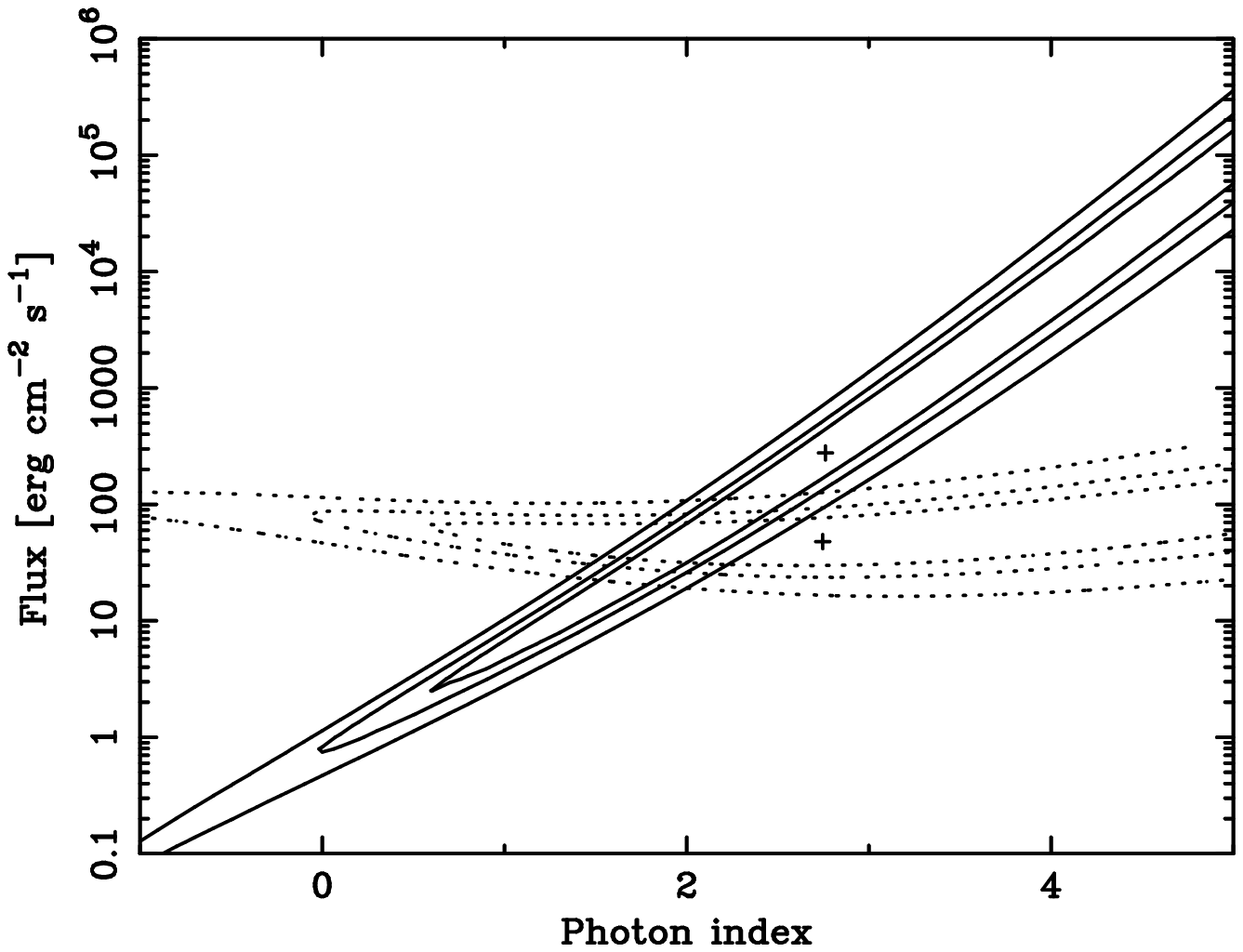}
\caption{
Two-parameter 68\%, 90\% and 99\% confidence contours of the absorbed power-law component. Only the photoelectric absorption is considered.
(a) Confidence contours of the photon index and absorption column density.
The dotted and solid contours correspond to
the result of the PDS fit (Model c)
and the combined fit (Model e), respectively.
(b) Confidence contours of the photon index and absorption-corrected flux
determined by the combined fit of the MECS and PDS (Model e).
The solid and dotted contours correspond to
the 2--10 keV flux and 20--100 keV flux, respectively.
\label{fig4}
}
\end{figure}

\begin{figure}
\epsscale{1.0}
\plotone{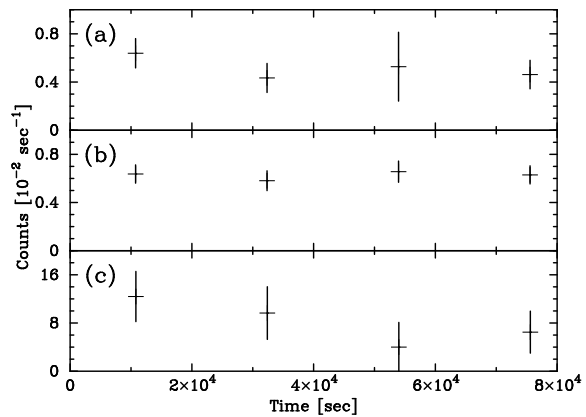}
\caption{
The LECS (a), MECS2+3 (b) and PDS (c) light curves of NGC 3079.
\label{fig5}
}
\end{figure}

\clearpage

\begin{deluxetable}{llllllll}
\tabletypesize{\scriptsize}

\tablecaption{Best-fit parameters of the NGC~3097 spectra. \label{tbl-1}}
\tablewidth{0pt}
\tablehead{
\colhead{Model}&&
	\colhead{(a)}&
		\colhead{(b)}&
			\colhead{(c)}&
				\colhead{(d)}&
					\colhead{(e)}&
						\colhead{(f)}\\
\colhead{Detector}&&
	\colhead{MECS}&
		\colhead{MECS}&
			\colhead{PDS}&
				\colhead{PDS}&
					\colhead{MECS+PDS}&
						\colhead{MECS+PDS}
}
\startdata
Power-law
&Photon index&
	$2.04^{+0.87}_{-0.67}$& 
		$1.74^{+0.55}_{-0.49}$& 
			---&                       
				---&                       
					$1.74^{+0.55}_{-0.49}$&    
						$1.75^{+0.55}_{-0.48}$\\
&2--10 keV flux\tablenotemark{a}&
	$3.3^{+1.1}_{-1.0}$& 
		$3.7\pm 0.8$& 
			---&                       
				---&                       
					$3.7\pm0.8$&    
						$3.7\pm0.8$\\
Power-law	
&Photon index&
	---&			
		 ---&                    
			$2.70^{+4.60}_{-2.62}$&    
				1.9 (fix)&                 
					$2.75^{+4.72}_{-2.31}$&    
						1.9 (fix)\\
&2--10 keV flux\tablenotemark{b}&
	---&			 
		---&                    
			$4.9^{+4.8}_{-2.0}$& 
				$3.8^{+2.0}_{-1.5}$& 
					$4.9^{+6.0}_{-2.1}$& 
						$3.8^{+2.0}_{-1.4}$\\
&20--100 keV flux\tablenotemark{c}&
	---&			 
		---&                    
			---&                       
				$4.8^{+2.4}_{-1.9}$& 
					---&                       
						$4.8^{+2.5}_{-1.7}$\\
&Absorption\tablenotemark{d}&
	---&			 
		---&                    
			$15.2^{+42.5}_{-15.2}$&    
				$9.6^{+11.7}_{-6.6}$&      
					$15.8^{+46.0}_{-12.7}$&    
						$10.0^{+11.4}_{-5.3}$\\
Gaussian	
&Energy	[keV]&
	$6.41^{+0.28}_{-0.33}$& 
		$6.39^{+0.28}_{-0.16}$& 
			---&                       
				---&                       
					$6.39^{+0.28}_{-0.17}$&    
						$6.39^{+0.28}_{-0.17}$\\
&Line width [keV]&
	$0.44^{+0.56}_{-0.44}$& 
		0 (fix)&                
			---&                       
				---&                       
					0 (fix)&                   
						0 (fix)\\            
&Normalization\tablenotemark{e}&
	$12.4^{+8.6}_{-7.6}$&	 
		$8.0\pm 4.1$&           
			---&                       
				---&                       
					$8.0^{+4.1}_{-4.0}$&       
						$8.0\pm 4.1$\\       
\tableline 
$\chi^2$/d.o.f.&&
	0.86/4&
		1.98/5&
			0.10/1&
				0.29/2&
					2.08/6&
						2.30/7\\
&&	0.215& 
		0.396& 
			0.098& 
				0.145& 
					0.347& 
						0.328\\
\enddata
\tablenotetext{a}{In unit of 10$^{-13}$ erg s$^{-1}$ cm$^{-2}$.}
\tablenotetext{b}{In unit of 10$^{-11}$ erg s$^{-1}$ cm$^{-2}$.}
\tablenotetext{c}{In unit of 10$^{-11}$ erg s$^{-1}$ cm$^{-2}$. Shown after absorption correction.}
\tablenotetext{d}{Hydrogen column density $N_{\rm H}$ in unit of 10$^{24}$ cm$^{-2}$. Only the photoelectric absorption is considered.}
\tablenotetext{e}{Photons in the line in unit of 10$^{-6}$ cm$^{-2}$ s$^{-1}$.}

\tablecomments{Errors refer to single-parameter 90\% confidence limits.
The PDS/MECS normalization factor is fixed at 0.86.}
\end{deluxetable}
\end{document}